 \documentclass[aps,prl,preprint,superscriptaddress,showpacs]{revtex4}
\usepackage{graphicx}

\begin{document}
\newcommand{\pH}{p-H$_2$}
\newcommand{\oD}{o-D$_2$}
\bibliographystyle{apsrev}


\title{Diffusion Monte Carlo study of the equation of state of solid ortho-D$_2$}



\author{Francesco Operetto}

\affiliation{Dipartimento di Fisica and INFN, Universit\`a di Trento, via Sommarive, 14
I-38050 Povo, Trento, Italy
}

\author{Francesco Pederiva}
\affiliation{Dipartimento di Fisica and INFN, Universit\`a di Trento, via Sommarive, 14
I-38050 Povo, Trento, Italy
}
\affiliation{CNR/INFM DEMOCRITOS National Simulation Center, Trieste, Italy}


\date{\today}

\begin{abstract}
We present results of Diffusion Monte Carlo calculations for a system of solid 
ortho-D$_2$ at different densities, for pressure ranging from 0 up to 350MPa. We 
compare the equation of state obtained using two of the most used effective intermolecular 
potentials, {\em i. e.} the Silvera--Goldman and the Buck potentials, with experimental data, 
in order to assess  the validity of the model interactions. The Silvera-Goldman potential 
has been found to provide a satisfactory agreement with experimental results,  
showing that, as opposed to what recently found for p-H$_2$, three--body forces
can be efficiently accounted for by an effective two--body term.
\end{abstract}

\maketitle


\section{Introduction}
Solid molecular hydrogen
has been the subject of intense experimental and theoretical study over the years.
However, the properties
of bulk isotopic equivalents (like HD or D$_2$) were not investigated in comparable 
detail.    

It is a common practice to describe molecules of para-hydrogen and isotopic equivalents as
point particles interacting via a model potential which in some case includes 
terms which effectively account for three--body interactions, which are expected
to be not negligible in these systems. 
The popular Silvera--Goldman (SG)\cite{Sil78} potential, which includes a $1/r^9$ term 
which mimicks a triple--dipole interaction, and the 
Buck\cite{Buc83,Nor84} potential, which is instead a pure two--body interaction, 
have been employed indifferently in theoretical works
about \pH, while in the case of \oD\ only the first one has been commonly used.
However, in the case of \pH, we recently discovered by means of accurate Quantum Monte
Carlo simulations that an explicit three-body  term is necessary in order to 
correctly reproduce experimental data on pressure\cite{Ope06_eos_h2}, . 
It is therefore interesting to investigate if the different mass  plays a role
in determining the necessity of including an explicit three-body force in the 
description of \oD.  
As already mentioned, both the Buck and the SG interactions are isotropic and therefore either completely 
neglect, or treat in an effective way the possible three body interactions acting among the molecules. 
This commonly used approximation is in principle justified by the fact that at T=0K the molecules of hydrogen and deuterium 
are considered to be in their rotational ground-state (J=0) and therefore the average
interaction must have spherical symmetry. Three body interactions should therefore be
a second order effect.

In this work we present the results of a systematic study of the equation of state
of solid \oD\ for pressures ranging from 0 (and slightly below) up to 350MPa 
performed by means of the Diffusion Monte Carlo algorithm, 
which gives for many Bosons the exact eigenvalue of the system, depending only 
on the chosen Hamiltonian.
We compared the results of our simulations with the experimental measurements
of the pressure. Only the SG potential is  found to be capable to give a satisfactory
description of the pressure--volume curve in the studied range of density. We can 
safely attribute this result to the presence in the potential of the repulsive extra-term 
which treats in an effective way the possible three body interactions acting among 
the molecules.

The next section of the paper will describe the methods used in our analysis.
Section III will present the results on energies, pressures and and other ground state properties
in \oD. Section IV is devoted to conclusions.  

\section{Methods}
We model a system of solid \oD\ as a set of $N$\ point particles described  by the following Hamiltonian:
\begin{equation}
\mathcal{H} = -\frac{\hbar^2}{2m}\sum_{i=1}^{N}\nabla^2_i 
+ \sum_{i<j} v_2(r_{ij}), 
\end{equation}
where $R=\{\bf r_1\dots \bf r_N\}$\ stands for the $3N$\ coordinates of the molecules 
and $v_2(R)$\ is the intermolecular potential. As already mentioned, we use two different molecule--molecule 
interaction $v_2$, i.e. the Silvera--Goldman and the Buck potentials.

Diffusion Monte Carlo is a stochastic algorithm projecting the ground state of a many--body Bosonic system 
by starting from a variational ansatz for the many--body wavefunction $\Psi_T(R)$
and evolving the function
$f(R,\tau)=\Psi_T(R)\Psi(R,\tau)$\ in imaginary time $\tau=it$\ according
to the time--dependent Schr\"odinger equation:
{\setlength\arraycolsep{2pt}
\begin{eqnarray}\label{eq:dmc}
-\frac{\partial f({R},\tau)}{\partial\tau}
&=&-\frac{1}{2}\mathbf{\nabla}^2f(R,\tau)+{\nabla}\cdot
\Big[\mathbf{F}(R)f(R,\tau)\Big]\nonumber\\
&\phantom{\ }&+\Big[E_L(R)-E_T\Big]f(R,\tau).
\end{eqnarray}}
Here the \emph{quantum force} ${F}$\ is defined as:
\begin{equation}
\mathbf{F}(R)=\mathbf{\nabla}\ln\left[\Psi_T(R)\right]=\frac{\mathbf{\nabla}\Psi_T(R)}{\Psi_T(R)}
\end{equation}
and the coefficient of the \emph{branching} term is defined in terms of
the \emph{local energy}:
\begin{equation}
E_L({R})=\frac{\mathcal{H}\psi_T({R})}{\Psi_T(R)}.
\end{equation}
Under the assumption that 
\begin{equation}
\langle\Psi({R},0)\vert\phi_0\rangle=\langle\Psi_T({R})\vert\phi_0\rangle\neq0,
\end{equation}
in the limit of $\tau\to+\infty$\ the distribution $f({R},\tau)$\ becomes proportional
to $\Psi_T({R})\phi_0({R})$.

In the actual simulation one evolves a population of $N_w$ points in configuration space ({\it walkers})
drawn from the initial distribution $f(R,0)=\vert\Psi_T\vert^2$
according the approximate importance sampled Green's function (for details about the sampling algorithm see 
{\em e. g.} Ref. \onlinecite{Sar02}):
\begin{equation}
G({\bf R},{\bf R'},\Delta\tau)=G_0({\bf R},{\bf R'},\Delta\tau)\frac{\psi({\bf R'})}{\psi({\bf R})}
\end{equation}
where
\begin{equation}
G_0({\bf R},{\bf R'},\Delta\tau)e^{-\frac{\Delta\tau}{2}V({\bf R})}
e^{\frac{({\bf R}-{\bf R'})^2}{2D\Delta\tau}}
e^{-\frac{\Delta\tau}{2}V({\bf R'})}.
\end{equation}
The propagation has to be iterated in order to achieve a propagation for an imaginary time
$\tau$ such that the walkers evolved in the Monte Carlo are asymptotically sampled from
$\Psi_T\phi_0$
In this situation the expectation value of the 
Hamiltonian itself computed as the average of the local energy $\mathcal{H}\psi(R)/\psi(R)$ on the 
sampled walkers becomes:
\begin{equation}
\langle \mathcal{H}\rangle = \frac{\int dR \Psi_T(R)\phi_0(R)\mathcal{H}\Psi_T(R)/\Psi_T(R)}
{\int dR \Psi_T(R)\phi_0(R)} = E_0,
\end{equation}
the lowest eigenvalue of the Hamiltonian. The computed estimate is exact at order $\Delta\tau$ 
and the results are are largely independent of the choice of the trial wave function.
Nevertheless, $\Psi_T$ has be variationally optimized, in order to reduce the variance 
on the estimate of the eigenvalue in the DMC calculation. 
We chosen the popular Jastrow-Nosanow form
\begin{equation}
\psi_T(R) = \prod_{i<j}\exp\bigg[\frac{1}{2}u(r_{ij})\bigg]\prod_{i=1}^N\exp\Big[-C(\bf r_i - \bf S_i)^2\Big],
\label{wavefn}
\end{equation}
where the $\{\bf S_i\}$ are the coordinates of the lattice sites around which the
molecules are confined.

The possibility of neglecting the symmetrization under exchange of molecules
in the trial wavefunction has been already discussed in the case of solid \pH\cite{Ope06_eos_h2},
justifing this ansatz also for solid \oD\ which is expected to be even more localized than and \pH.

A series expansion of the Jastrow factor in equation (\ref{wavefn}) was used
in order to improve the description of the two--body correlations in the system.
The function $u(r)$ is the sum of a standard McMillan pseudopotential
and of a correction term which has been expanded in terms of suitable basis functions:
\begin{equation}
u(r)=\left(\frac{b}{r}\right)^5+\sum_n a_n\chi_n(r).
\end{equation}
The basis functions $\chi_n$ (about 40) are the same used in Ref. \onlinecite{Ope06_eos_h2}
for analogous calculations in solid \pH:
\begin{equation}
\chi_n(r)=\left\{\begin{array}{ll}
\Big\{1-\cos\Big[\frac{2\pi n}{L/2-r_c}(r-L/2)\Big]\Big\}r^{-5} & 
\textrm{\ } r>r_c\\
0 & \textrm{\ } r\le r_c \end{array}.\right.
\end{equation}
The cutoff radius $r_c$ allows to remove  divergences for $r\to0$ and therefore to 
avoid numerical instabilities in the optimization procedure.
If such cutoff is small enough, it does not influence the value of the energy.
In our calculations it has been taken equal to 1\AA. 
The parameters $b$, $C$ and $\{a_n\}$ have been determined by following a reweighting scheme,
which provides to alternate between VMC calculations of $\langle \mathcal{H}\rangle$
and minimization of a combination of the variance of the expectation value of the 
Hamiltonian, and of the expectation value itself on a set of sampled configurations.

For each density studied, DMC calculations have been performed projecting from the optimized 
trial function $\Psi_T$ using  different imaginary time-steps
and populations of walkers, 
in order to check that the bias due to the use of a finite time-step and a limited number of walkers were both 
negligible compared to statistical errors affecting the results of the simulations. 
As well as the ground state energy, the configuration generated in the QMC simulations asymptotically sampling
from $\Psi_T\phi_0$ can be used to evaluate the mixed estimator
\begin{equation}
\langle\mathcal{O}\rangle_{\textrm{\tiny mix}}=
\frac{\langle\Psi_T\vert\mathcal{O}\vert\phi_0\rangle}{\langle\Psi_T\vert\phi_0\rangle} 
\simeq\frac{1}{N_w}\sum_{i=1}^{N_w}\frac{\mathcal{O}\Psi_T(R_i)}{\Psi_T(R_i)}
\end{equation}
of any operator $\mathcal{O}$ of interest.
The mixed estimator coincides with the ground-state expectation value 
$\langle\phi_0\vert\mathcal{O}\vert\phi_0\rangle$
only if $\mathcal{O}$ commutes with the Hamiltonian, otherwise one can obtain an estimate 
of the  ground-state expectation by computing the extrapolated estimator 
$\langle\mathcal{O}\rangle_{\textrm{\tiny ext}}=
2\langle\mathcal{O}\rangle_{\textrm{\tiny mix}}-
\langle\Psi_T\vert\mathcal{O}\vert\Psi_T\rangle/\langle\Psi_T\vert\Psi_T\rangle$
The bias affecting such estimate is of the second order in $\alpha$, where
$\phi_0=\Psi_T+\alpha\Psi_\alpha$.

We performed simulations at different densities ($0.0194\textrm{\AA}^{-3}<\rho<0.0430\textrm{\AA}^{-3}$)
for two different lattices, face centered cubic (fcc) and hexagonal close packed (hcp). Simulation cell were set to 
accommodate $3\!\times\!3\!\times\!3$ elementary cubic cells for fcc, with a total of $N$=108 lattice sites, and 
$5\!\times\!3\!\times\!3$ elementary cells for 
a total of $N$=180 lattice sites for hcp.
Periodic boundary conditions are also imposed in order to reduce finite--size effects.
The molecule--molecule interaction is truncated at the edge of a sphere of radius equal to 
$L/2$, where $L$ is the length of the shortest side of the cell. In
order to avoid discontinuities in the Green's function used to sample the walkers,
the potential was also shifted of a quantity $v(L/2)$. 
The contribution from the potential energy outside the sphere is estimated 
by integrating the potential in the $(L/2,+\infty]$ interval, assuming
therefore a uniform pair correlation function of molecules beyond $L/2$. A similar 
procedure has been used for correcting the error due to the shift in the sphere of radius $L/2$. 

\section{Results}  
\begin{figure}
\begin{center}
\includegraphics[scale=0.32]{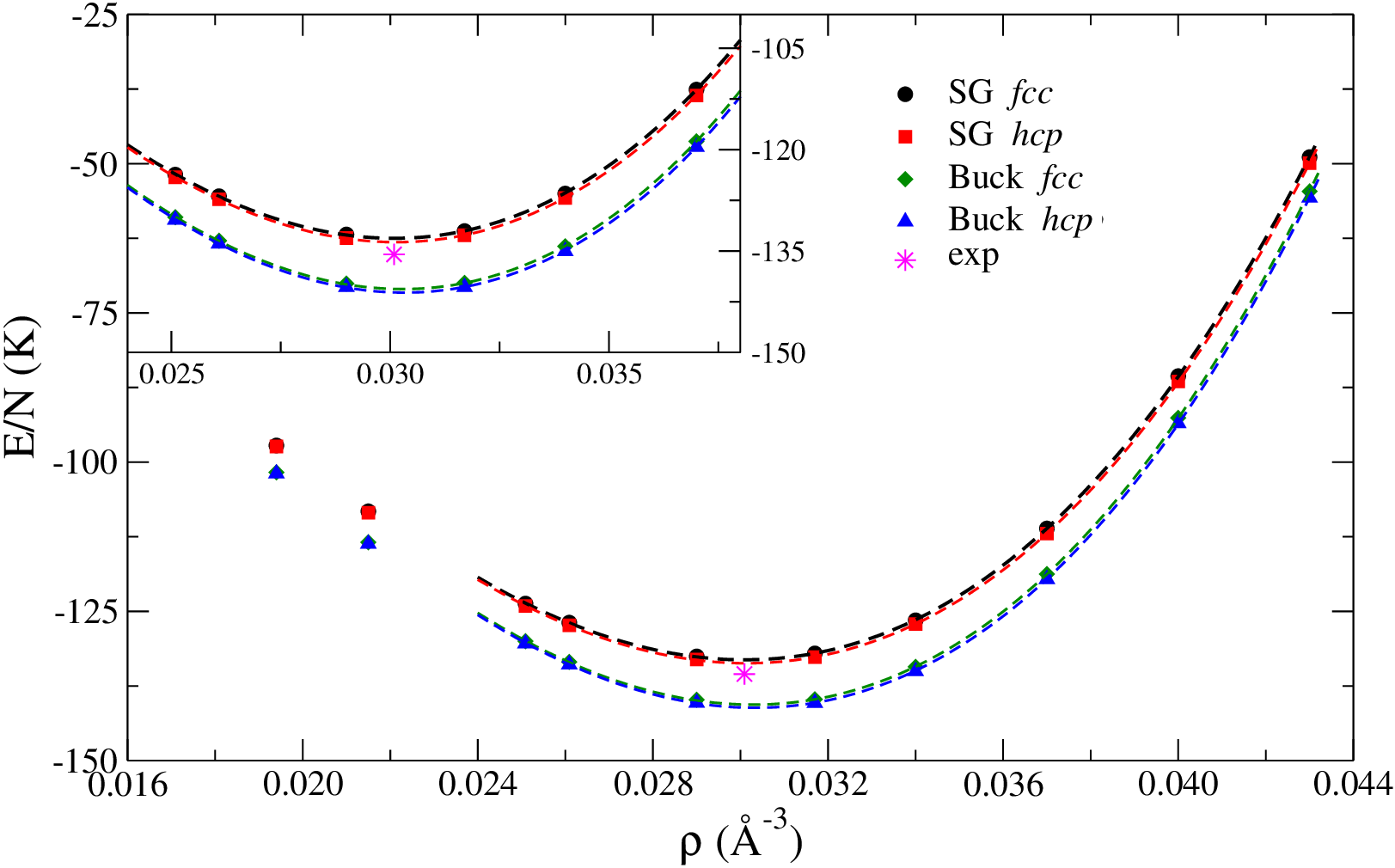}
\caption{Energy per particle (in K) in solid \pH\ as a function of the density $\rho$,
computed with different interactions. The curves represent the fits to the DMC
results with Murnaghan--like curve. In the inset the region around the equilibrium
density is expanded. The experimental point is taken from Ref. \onlinecite{Sch70}.} 
\label{fig:en}
\end{center}
\end{figure}

In Fig. \ref{fig:en} we report the results obtained for the equation of state in the density interval
$0.0194$\AA$^{-3}<\rho<0.0430$\AA$^{-3}$, for the SG and the Buck potentials. In the inset we 
report the expanded curves for density near the equilibrium one. 
The curves report the DMC results, which are found to be only slightly lower than the
corresponding results obtained by optimizing the Jastrow factor in the wavefunction.
We found that the difference between VMC and DMC results is less than 2.5K in all the range of density studied 
and depends only weakly on the density.
This suggests that the trial wavefunction used is already quite accurate. 

\begin{table}
\begin{tabular}{c|c|c|c|c}
\toprule
$\rho$  &SG fcc&SG hcp&Buck fcc&Buck hcp\\
\hline
0.01940 &  -97.21(1) &  -97.37(3) & -101.73(2) & -101.91(2)\\
0.02150 & -108.23(2) & -108.48(4) & -113.42(2) & -113.69(1)\\
0.02509 & -123.70(2) & -124.09(2) & -129.99(1) & -130.38(2)\\
0.02609 & -126.89(2) & -127.33(2) & -133.47(1) & -133.90(1)\\ 
0.02900 & -132.55(2) & -133.08(2) & -139.81(2) & -140.33(1)\\
0.03170 & -132.05(2) & -132.67(2) & -139.73(1) & -140.35(1)\\
0.03400 & -126.49(3) & -127.14(1) & -134.30(1) & -135.03(2)\\
0.03700 & -111.11(3) & -111.97(2) & -118.77(1) & -119.62(2)\\
0.04000 &  -85.59(3) &  -86.50(3) &  -92.58(1) &  -93.50(2)\\
0.04300 &  -48.89(4) &  -49.92(3) &  -54.63(3) &  -55.71(2)\\ 
\botrule
\end{tabular}
\caption{Energy per particle (in K) in solid \oD\ computed with different 
interaction for fcc and hcp crystalline structure.}
\label{tab:en}
\end{table}
\begin{table}
\begin{tabular}{c|c|c|c|c}
\toprule
$\rho$  &108 (SG)&256 (SG)&108 (Buck)&256 (Buck)\\
\hline
0.02609 & -126.89(2) & -126.77(1) & -133.47(1) & -133.35(1) \\
0.04300 &  -48.89(4) &  -48.52(2) & -54.63(3)  & -54.33(3)  \\
\botrule
\end{tabular}
\caption{Energy per particle (in K) in solid \oD\ computed in
fcc crystal using different number of molecules interacting via the Buck potential. 
Densities are in \AA${}^{-3}$.}
\label{tab:256}
\end{table}
We fitted the computed DMC results of the energy per particle (Table \ref{tab:en}) by means 
of a modied Murnaghan curve:
\begin{equation}
\label{eq:eos}
\epsilon(\rho)=a+b\rho+c\rho^\gamma.
\end{equation}
The coefficients and the minimum of the curves in the fcc and hcp crystal are reported
in Table \ref{tab:coeff}. 
\begin{table}
\begin{tabular}{l|c|c|c|c|c|c}
\toprule
\phantom{Buck fcc} & $a$ & $b$ & $c$ & $\gamma$ & $\rho_0$ & $\epsilon_{0}$ \\
\hline
SG fcc      & 80.4313 & -9932.81 & 1.81514$\cdot$10${}^{7}$ & 3.50150 & 0.03009(1) & -133.12(4)\\
SG hcp      & 79.9612 & -9912.63 & 1.86138$\cdot$10${}^{7}$ & 3.51105 & 0.03014(1) & -133.69(4)\\
Buck fcc    & 85.0026 & -10397.5 & 2.04201$\cdot$10${}^{7}$ & 3.52878 & 0.03028(1) & -140.63(4)\\
Buck hcp    & 83.5366 & -10313.4 & 2.16255$\cdot$10${}^{7}$ & 3.55035 & 0.03033(1) & -141.14(4)\\
\botrule
\end{tabular}
\caption{Coefficients of the equation of state (Eq. $\ref{eq:eos}$)
fitted to DMC simulation results and computed equilibrium energy and equlibrium density. 
The equilibrium energies $\epsilon_0$ are given in Kelvin;
the equilibrium densities $\rho_0$ are given in \AA${}^{-3}$.
Units of the parameters $a$, $b$ and $c$ are expressed accordantly.}
\label{tab:coeff}
\end{table}
The only available experimental estimate of the sublimation heat at 0K for solid D$_2$ is
-135.5K per molecule\cite{Sch70} and refers to solid normal D$_2$ (33\% p-D$_2$), whose binding energy is expected
to be slightly lower than that of \oD. The experimental value can be therefore considered only a lower bound 
for the binding energy of pure \oD.
In the case of molecular hydrogen, the difference in energy between normal 
and pure \pH\ is less than 4\%\cite{Nor84,Sch70}. 
The computed DMC values for \oD\ at equilibrium density of \oD\ using the Buck potential
($E/N= -140.63$K and $E/N=-141.14$K for the fcc and hcp crystal respectively) are found to be about 5K
below the experimental result. Such result confirms that employing an effective potential which completely 
neglects tree-body interactions, like the Buck one, leads to an overestimate of the binding energy of
the system, as found in the case of $^4$He\cite{Uje03} and \pH\cite{Ope06_eos_h2}.
On the other hand the SG potential gives results about 2K above the experimental lower bound.
This discrepancy is due to the effective
triple--dipole term proportional to $1/r^9$, which is repulsive, and which therefore tends to
increase the energy per particle. 
If we suppose that this effective term tends to overestimate the equilibrium energy as in the case 
of \pH, we can estimate the difference between the value found for the GS potential and the right value 
to be less than 2\%.

In both the cases the hcp lattice turns out to be always stable with respect to the fcc one. 
The equilibrium density computed by means of the SG potential has been found in better agreement with experimental 
finding $\rho_0=0.03009\textrm{\AA}^{-3}$ of Ref. \onlinecite{Dri79}.

In Tab. \ref{tab:256} we also reported some results of simulations performed in the fcc crystal
using 256 molecules arranged on $4\!\times\!4\!\times\!4$ elementary cells. 
The bias affecting the estimates of the energy due to the finite size of the simulation box have been found 
negligible in comparison with the energetic differences provided by the choice of different potentials. 
The energy gap 
between fcc and hcp lattice seems to slightly increase if the fcc lattice with 256 sites is taken into account.

\begin{figure}
\begin{center}
\includegraphics[scale=0.32]{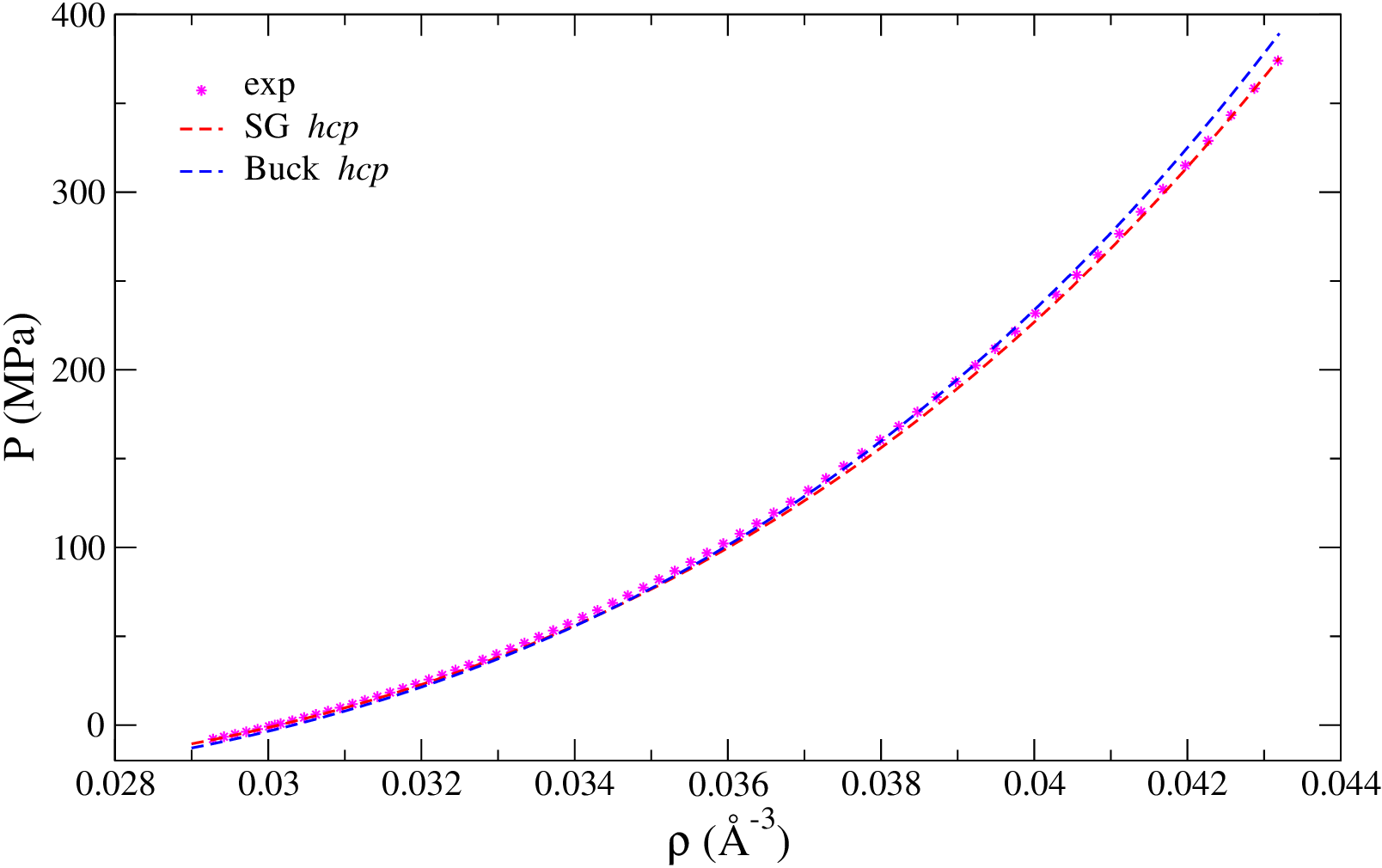}
\caption{Pressure as a function of the density $\rho$ in solid \pH\ computed 
with different interactions. Curves for fcc and hcp are almost indistinguishable. 
The experimental points are taken from 
Ref.\onlinecite{Dri79}}
\label{fig:press}
\end{center}
\vspace{-0.5cm}
\end{figure}

The pressure as a function of the density was then obtained from the fitted curves by means of 
the following expression:
\begin{equation}
P(\rho)=\rho^2\bigg[\frac{\partial\epsilon(\rho)}{\partial\rho}\bigg]_T.
\end{equation}
The computed curves for the pressure are reported in Fig. \ref{fig:press}, and compared with the experimental results of
Ref. \onlinecite{Dri79} for pure \oD. 
At low densities the curves obtained the two potentials tend to overlap and provide only a slight underestimation 
of the experimental pressure.
For densities $\rho > 0.037$\AA${}^{-3}$ the curve given by the Buck potential tends clearly to overestimate
the pressure, while the SG potential seems to be capable to keep a much better agreement with experimental data,
although the density dependence of the pressure indicates that the pressure would tend to be overestimated
at higher densities. It can be expected that at high densities the effects of many body terms in the potential 
become more important, but the the triple--dipole effective term in the SG potential seems to be capable to take into
account most of the many-body effects on the pressure curve.

The energy gap between the Buck and SG potential can be therefore considered as a first estimate of the 
contribution of three body terms to the potential, 
That contribution, ranging from 3 to 7K, represents about the 4\% of the absolute vaule 
of the two--body potential energy in all the range of studied density. In the case 
of solid \pH, indeed, the contribution of the three body term, although comparable, 
represents from 4\% up to the 12\% of the two--body potential energy.
depending on the density, leading to much stronger effects on the P--V curve, which have to 
be kept into account by adding an explicit Axilrod--Teller three--body term to the potential\cite{Ope06_eos_h2}.
The smaller relative influence of the many--body terms on the equation of state of \oD\ is to attribute 
to its heavier mass, which tends to keep the molecules more localized and therefore to increase
the two--body potential energy. 

The many--body effects in \oD\ are stronger than of $^4$He, 
although they have the same mass: the energetic contribution of
the three--body potential in $^4$He was estimated to be less than 1.2\% of the 
absolute value of the two--body potential energy at densities from 0.02934\AA$^{-3}$ to 0.03527\AA$^{-3}$.

Once the equation of state $e(\rho)$ is known, it is possible to calculate the isothermal compressibility,
defined as:
\begin{equation}\label{eq:k}
\kappa(\rho)=\frac{1}{\rho}\left[\frac{\partial\rho}{\partial P}\right]_T.
\end{equation}
\begin{figure}
\begin{center}
\includegraphics[scale=0.32]{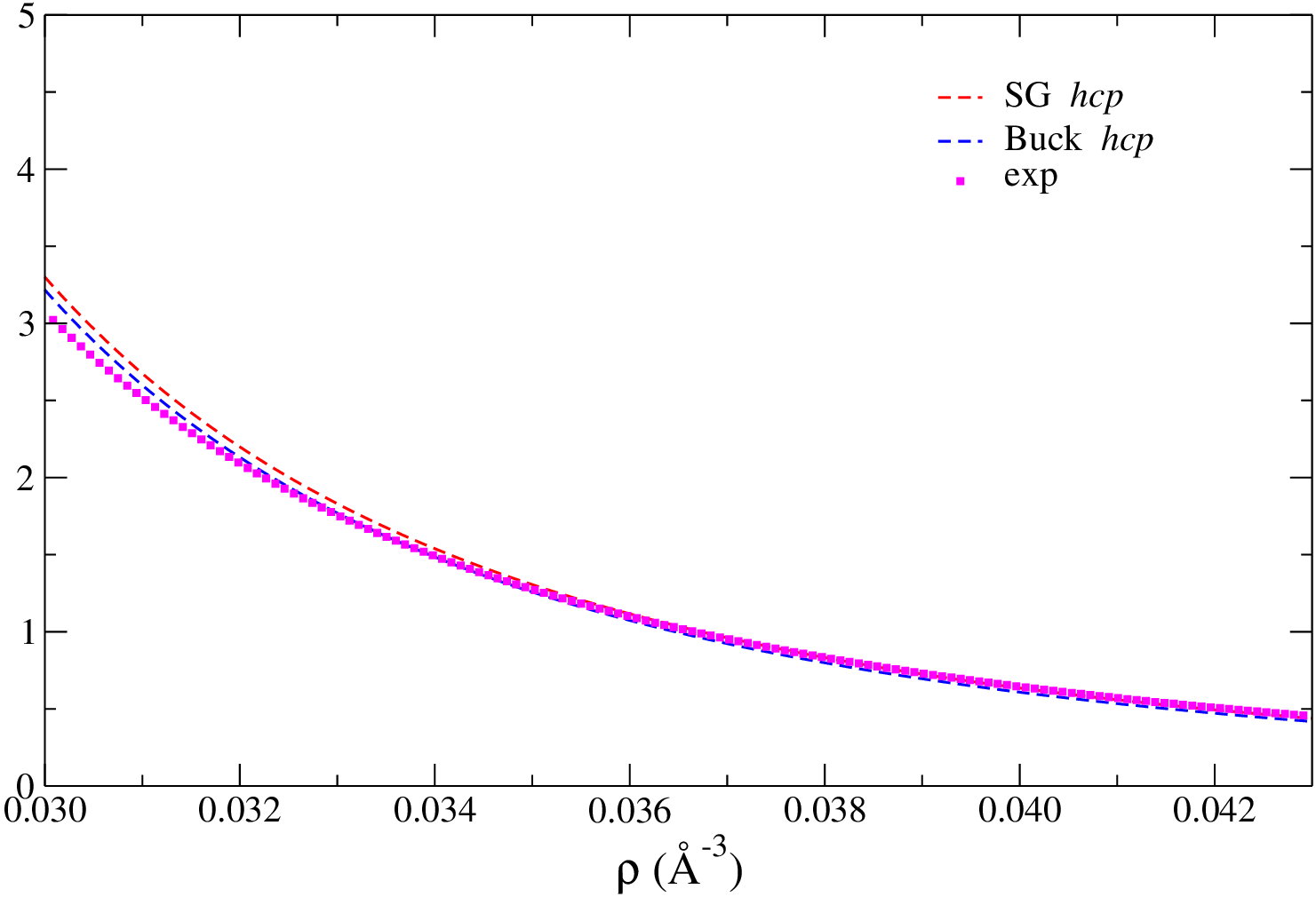}
\end{center}
\vspace{-0.5cm}
\caption{Compressibility in molecular \pH\ computed using the  SG and the Buck interactions. 
Curves for fcc and hcp are almost indistinguishable}
\label{fig:comp}
\end{figure}
In Fig. \ref{fig:comp} the results for the compressibility yelded by the two potentials
are compared with the corresponding experimental value obtained by fitting the pressure curve of 
Ref. \onlinecite{Dri79}. Both potentials provide a rather good agreement with experimental data except 
at the lowest densities, as expected due the bigger relative discrepancy between experimental and
calculated values of the pressure that arises when the pressure tends to vanish.
In order to investigate the effects of the choice of different models of intermolecular potentials
on the local structure of the \oD\ solid we computed the pair distribution function:
\begin{equation}
g(r)=\Bigg\langle\sum_{i\neq j}\delta(r_{ij}-r)\Bigg\rangle.
\end{equation}
\begin{figure}
\begin{center}
\includegraphics[scale=0.30]{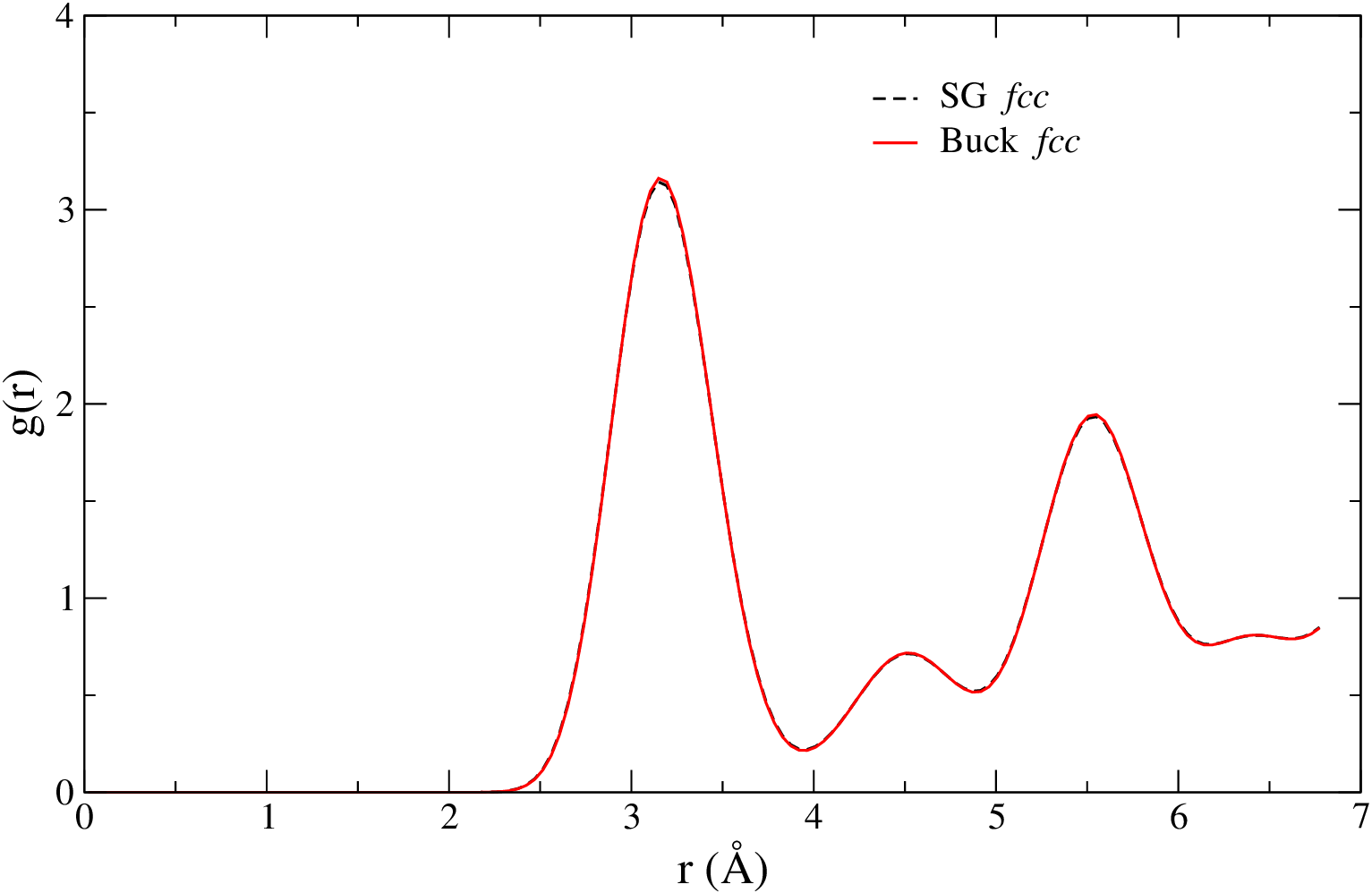}\hspace{0.3cm}
\includegraphics[scale=0.30]{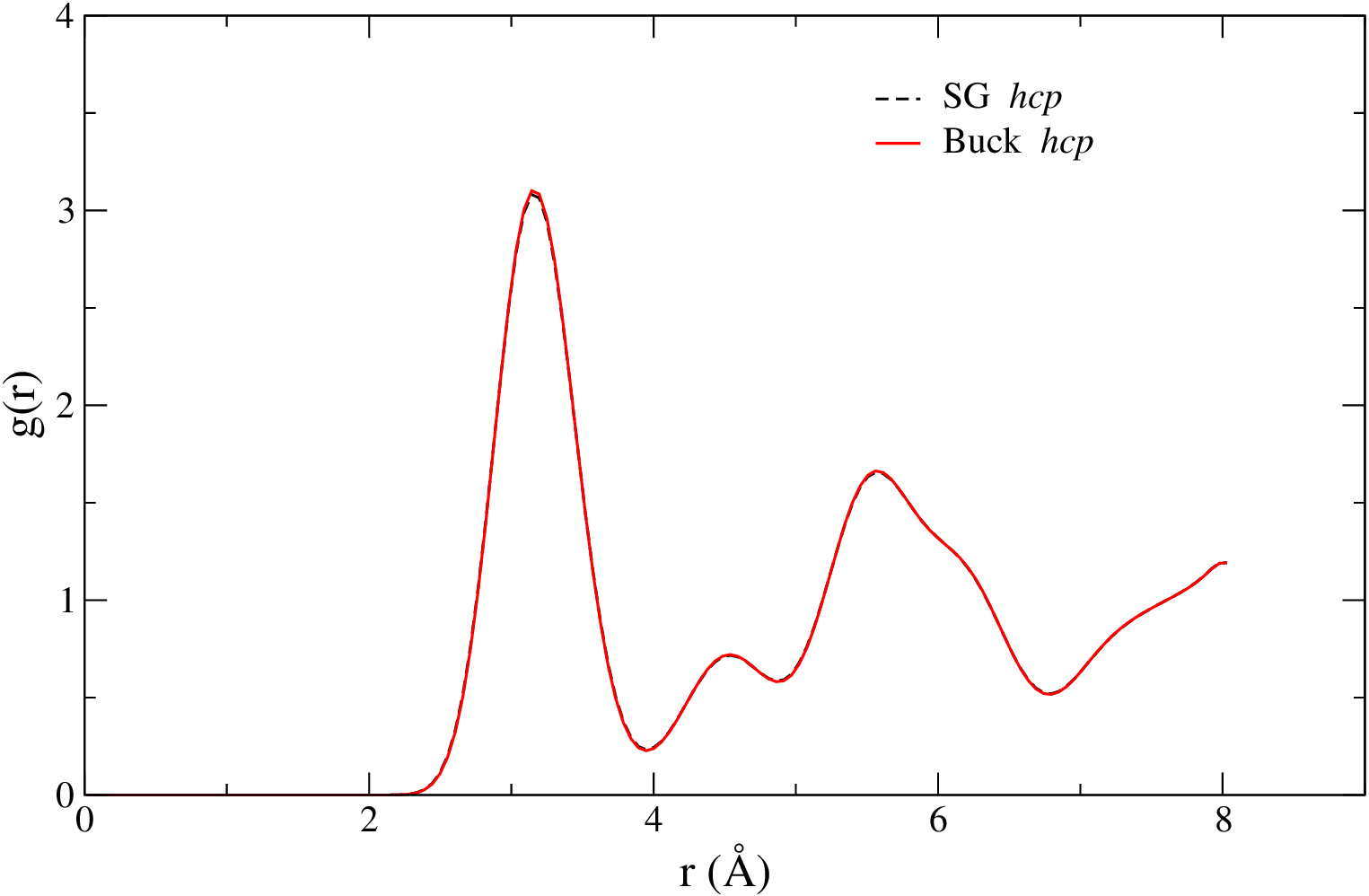}
\vspace{-0.5cm}
\caption{Pair distribution function in solid \pH computed using the SG
and the Buck potentials at density $\rho$=0.04300\AA$^{-3}$
both in the fcc and in the hcp crystal.}
\label{fig:gr}
\end{center}
\vspace{-0.5cm}
\end{figure}
Results for the computation of the function $g(r)$ are reported in Fig. \ref{fig:gr}. 
For all the range of density studied the curves obtained employing
the SG and the Buck potential are almost indistinguishable from each other,
giving an indication that the effects of the choice of different models of interaction
on the structural properties of the crystal of \oD\ are negligible, as already found 
also for solid \pH. 

\section{Conclusions}
We
present the results of a systematic comparison of ground state properties
of solid \oD\ computed with the two most used interactions, the Buck and the
Silvera--Goldman potential. Only the last one is capable to give a satisfactory
description of the pressure--volume curve. We attribute this result to the presence 
in the potential of a repulsive extra-term which treats in an effective way the 
possible three body interactions acting among the molecules. 

No evident structural
differences have been found by changing interaction model.
\section{Acknowledgments}
We are grateful to C.\,J.\,Umrigar and P.\,Nightingale for providing us 
the Levemberg-Marquardt minimization code used for optimizing the parameters 
in the variational wave function.

Calculations have been performed on  the HPC facility BEN at ECT* in Trento.
%
%

\begin{thebibliography}{8}
\expandafter\ifx\csname natexlab\endcsname\relax\def\natexlab#1{#1}\fi
\expandafter\ifx\csname bibnamefont\endcsname\relax
  \def\bibnamefont#1{#1}\fi
\expandafter\ifx\csname bibfnamefont\endcsname\relax
  \def\bibfnamefont#1{#1}\fi
\expandafter\ifx\csname citenamefont\endcsname\relax
  \def\citenamefont#1{#1}\fi
\expandafter\ifx\csname url\endcsname\relax
  \def\url#1{\texttt{#1}}\fi
\expandafter\ifx\csname urlprefix\endcsname\relax\def\urlprefix{URL }\fi
\providecommand{\bibinfo}[2]{#2}
\providecommand{\eprint}[2][]{\url{#2}}

\bibitem[{\citenamefont{Silvera and Goldman}(1978)}]{Sil78}
\bibinfo{author}{\bibfnamefont{I.~F.} \bibnamefont{Silvera}} \bibnamefont{and}
  \bibinfo{author}{\bibfnamefont{V.~V.} \bibnamefont{Goldman}},
  \bibinfo{journal}{J. Chem. Phys.} \textbf{\bibinfo{volume}{69}},
  \bibinfo{pages}{4209} (\bibinfo{year}{1978}).

\bibitem[{\citenamefont{Buck et~al.}(1983)\citenamefont{Buck, Huisken,
  Kohlhase, Otten, and Schaefer}}]{Buc83}
\bibinfo{author}{\bibfnamefont{U.}~\bibnamefont{Buck}},
  \bibinfo{author}{\bibfnamefont{F.}~\bibnamefont{Huisken}},
  \bibinfo{author}{\bibfnamefont{A.}~\bibnamefont{Kohlhase}},
  \bibinfo{author}{\bibfnamefont{D.}~\bibnamefont{Otten}}, \bibnamefont{and}
  \bibinfo{author}{\bibfnamefont{J.}~\bibnamefont{Schaefer}},
  \bibinfo{journal}{J. Chem. Phys.} \textbf{\bibinfo{volume}{78}},
  \bibinfo{pages}{4439} (\bibinfo{year}{1983}).

\bibitem[{\citenamefont{Norman et~al.}(1984)\citenamefont{Norman, Watts, and
  Buck}}]{Nor84}
\bibinfo{author}{\bibfnamefont{M.~J.} \bibnamefont{Norman}},
  \bibinfo{author}{\bibfnamefont{R.~O.} \bibnamefont{Watts}}, \bibnamefont{and}
  \bibinfo{author}{\bibfnamefont{U.}~\bibnamefont{Buck}}, \bibinfo{journal}{J.
  Chem. Phys.} \textbf{\bibinfo{volume}{81}}, \bibinfo{pages}{3500}
  (\bibinfo{year}{1984}).

\bibitem[{\citenamefont{Operetto and Pederiva}(2006)}]{Ope06_eos_h2}
\bibinfo{author}{\bibfnamefont{F.}~\bibnamefont{Operetto}} \bibnamefont{and}
  \bibinfo{author}{\bibfnamefont{F.}~\bibnamefont{Pederiva}},
  \bibinfo{journal}{{Phys Rev. B}} \textbf{\bibinfo{volume}{73}},
  \bibinfo{pages}{184124} (\bibinfo{year}{2006}).

\bibitem[{\citenamefont{Sarsa et~al.}(2002)\citenamefont{Sarsa, Boronat, and
  Casulleras}}]{Sar02}
\bibinfo{author}{\bibfnamefont{A.}~\bibnamefont{Sarsa}},
  \bibinfo{author}{\bibfnamefont{J.}~\bibnamefont{Boronat}}, \bibnamefont{and}
  \bibinfo{author}{\bibfnamefont{J.}~\bibnamefont{Casulleras}},
  \bibinfo{journal}{J. Chem. Phys.} \textbf{\bibinfo{volume}{116}},
  \bibinfo{pages}{5956} (\bibinfo{year}{2002}).

\bibitem[{\citenamefont{Schnepp}(1970)}]{Sch70}
\bibinfo{author}{\bibfnamefont{O.}~\bibnamefont{Schnepp}},
  \bibinfo{journal}{Phys. Rev. A} \textbf{\bibinfo{volume}{2}},
  \bibinfo{pages}{2574} (\bibinfo{year}{1970}).

\bibitem[{\citenamefont{Ujevic and Vitiello}(2003)}]{Uje03}
\bibinfo{author}{\bibfnamefont{S.}~\bibnamefont{Ujevic}} \bibnamefont{and}
  \bibinfo{author}{\bibfnamefont{S.~A.} \bibnamefont{Vitiello}},
  \bibinfo{journal}{Phys. Rev. B} \textbf{\bibinfo{volume}{119}},
  \bibinfo{pages}{8482} (\bibinfo{year}{2003}).

\bibitem[{\citenamefont{Driessen et~al.}(1979)\citenamefont{Driessen, de~Waal,
  and Silvera}}]{Dri79}
\bibinfo{author}{\bibfnamefont{A.}~\bibnamefont{Driessen}},
  \bibinfo{author}{\bibfnamefont{J.~A.} \bibnamefont{de~Waal}},
  \bibnamefont{and} \bibinfo{author}{\bibfnamefont{I.~F.}
  \bibnamefont{Silvera}}, \bibinfo{journal}{J. Low Temp. Phys.}
  \textbf{\bibinfo{volume}{34}}, \bibinfo{pages}{255} (\bibinfo{year}{1979}).

\end{thebibliography}

\end{document}